\documentclass[%
 aip,
 amsmath,amssymb,
 reprint,%
]{revtex4-1}

\usepackage{graphicx}
\usepackage{dcolumn}
\usepackage{bm}

\usepackage[utf8]{inputenc}
\usepackage[T1]{fontenc}
\usepackage{mathptmx}
\usepackage{xcolor}
\usepackage{ulem}
\usepackage{hyperref}
\begin{document}

\preprint{AIP/123-QED}

\title{Topological transitions to Weyl states in bulk Bi$_2$Se$_3$: Effect of hydrostatic pressure and doping}
\author{Sudip Kumar Saha}
\affiliation{Department of Condensed Matter Physics $\&$ Material Sciences, \\S.N. Bose National Centre for Basic Sciences, \\JD Block, Sector-III, Salt Lake City, Kolkata 700 098, India \\ }
\author{Hrishit Banerjee}
\affiliation{Institute of Theoretical and Computational Physics, Graz University of Technology, NAWI Graz, Petersga{\ss}e 16, Graz, 8010, Austria.}
\email{h.banerjee10@gmail.com}
\author{Manoranjan Kumar}
\affiliation{Department of Condensed Matter Physics $\&$ Material Sciences, \\S.N. Bose National Centre for Basic Sciences, \\JD Block, Sector-III, Salt Lake City, Kolkata 700 098, India }
\email{manoranjan15@gmail.com}

\date{\today}
\begin{abstract}
Bi$_2$Se$_3$, a layered three dimensional (3D) material, exhibits topological insulating properties due to presence of surface states and a band gap of 0.3 eV in the bulk. We study the effect of hydrostatic pressure $P$ and doping with rare earth elements on the topological aspect of this material in bulk from a first principles perspective. Our study shows that under a moderate pressure of P$>$7.9 GPa, the bulk electronic properties show a transition from an insulating to a Weyl semi-metal state due to band inversion. This electronic topological transition may be correlated to a structural change from a layered van der Waals material to a 3D system observed at $P$=7.9 GPa. At large $P$ density of states have significant value at the Fermi-energy. Intercalating Gd with a small doping fraction between Bi$_2$Se$_3$ layers drives the system to a metallic anti-ferromagnetic state, with Weyl nodes below the Fermi-energy. At the Weyl nodes  time reversal symmetry is broken due to finite local field induced by large magnetic moments on Gd atoms. However, substituting Bi with Gd induces anti-ferromagnetic order with an increased direct band gap. Our study provides novel approaches to tune topological transitions, particularly in capturing the elusive Weyl semimetal states, in 3D topological materials.
\end{abstract}

\maketitle

\section{Introduction}
Topological insulators (TI) have potential future application in quantum computers\cite{Moore2009, Moore2010, nayak} and spintronics \cite{yokoyama, fan, sun} owing to existence of symmetry protected edge or surface states, and also provide a fundamental bridge between high-energy and condensed-matter physics due to the presence of exotic physical states in the system. These materials exhibit insulating bulk and metallic surface states and these properties lead to extensive theoretical and experimental studies \cite{Moore2010,Hasan2, Hasan, Kane, Chen,sanman}. These systems show non-trivial topological order by conserving the particle number and time reversal symmetry\cite{Frank, Chen}.

The primary feature of the TI state is the inverted band structure, which results from the crossing of the valence and conduction bands of different parity symmetry\cite{betancourt2016, zhang_2010}, and Bi$_x$Sb$_{1-x}$ family of materials are a prime example of three dimensional (3D) materials with $Z_2$ invariant symmetry. Sb$_2$Te$_3$, Bi$_2$Te$_3$ and Bi$_2$Se$_3$ are  3D material topological layered materials. Bi$_2$Se$_3$ forms effectively two dimensional (2D) layered structures and theoretical studies predict that Bi$_2$Se$_3$ has a topologically non-trivial bulk energy gap of 0.3 eV \cite{zhang,cava,Park_2016, zhang_2010,cava2,chiatti}. The topological surface states are described by a single gapless Dirac cone at the $\Gamma$ point \cite{zhang,zhang2,arijitsaha,cava2}. It is interesting to understand the mechanism which may close bulk energy gap in Bi$_2$Se$_3$.   

 Bi$_2$Se$_3$ and its family of materials have been extensively explored, by applying pressure \cite{waghmare,young,liu,polian,hamlin, vilaplana,gomis}, doping Bi with rare earth (RE) atoms  like Gd, Sm etc. \cite{song,zheng,kholdi,kim2,Li-apl,deng} and also with transition metal (TM) atoms like Cr, Fe, etc.\cite{choi-appl,choi-appl2,dyck, checkelsky}. Incorporating impurities with large magnetic moment may break time reversal symmetry and lead to many exotic phenomena like quantum anomalous hall effect (QAHE) which supports dissipationless charge transport. Mn-doped Bi$_2$Se$_3$ shows spin glass like behavior \cite{choi-appl}, whereas, 
 Fe and Cr-doping leads to a dominant of ferromagnetic and dominant anti-ferromagnetic interactions respectively \cite{choi-appl2}.  A small ( less than 0.1 \%) doping of Cr in this material are reported as ferromagnetic, while Fe-doped Bi$_2$Se$_3$ tends to be a weakly anti-ferromagnetic system. Cr and Fe-doped Bi$_2$Se$_3$ are insulating, but the band gaps are substantially reduced due to the strong hybridization between the $d$ orbitals of the dopants and the $p$ orbitals of the neighboring Se atoms \cite{abramson}.  
 
Doping of Bi$_2$Se$_3$  with TM atoms can lead to many exotic phases; for e.g. doping with Cu with doping fractions of 0.12 and 0.15 shows intermixing of both Cu-intercalation between Se-Se layers and Cu-substitution in Bi-layer sites \cite{cu-inter}. In Cu-intercalated Bi$_2$Se$_3$ superconducting transitions are  observed experimentally with a $T_C$ of 3.5 K and 3.6 K for doping fractions of 0.12 and 0.15 respectively. However, at low temperature Cu-doped Bi$_2$Se$_3$ crystals behaves like a paramagnetic metal \cite{cu-inter}. Intercalation of Cu in this material also holds the promise of topological superconductivity \cite{frohlich}.

RE atoms with $4f$ electrons  are  expected to be  better candidates for introducing magnetic order in TIs compared to $3d$ TM atoms \cite{deng, chang-prl}, because of their larger number of possible unpaired electrons, larger ionic radii compared to TM atoms,  as well as the fact that RE radii are comparable to Bi atoms of Bi$_2$Se$_3$.  Therefore, a smaller structural distortion is expected due to small atomic radii mismatch, and no disorder is envisioned. \cite{deng}. RE atoms are better dopants in avoiding impurity aggregation which is seen in Cr (TM) doped (Bi,Sb)$_2$Te$_3$ and these aggregations are responsible for magnetic disorder\cite{chang-prl}.  In addition $f$ orbitals can have more unpaired spins compared to $d$ orbitals, therefore RE can induce larger magnetic moments compared to TM atoms. $4f$ RE electrons are more localized with a maximum of 7 unpaired electrons compared to a maximum of 5 unpaired electrons in $3d$ TMs. In some cases a bulk paramagnetic behavior with a large magnetic moment from substituted Gd$^{3+}$ ion is reported in Bi$_2$Se$_3$ \cite{song}; whereas, a doping induced paramagnetic to anti-ferromagnetic phase transition is observed experimentally on substitutional doping of the system with Gd \cite{kim2}. Intercalating Rb atoms between the quintuple layer structure of Bi$_2$Se$_3$ can form a quantum-confined two-dimensional electron gas state (2DEG) with a strong Rashba-type spin–orbit splitting \cite{Bianchi2012}.


Application of pressure is an important tool to enhance the hybridization between the orbitals, and a recent study shows the emergence of an unconventional superconducting phase in topological Bi$_2$Se$_3$ at a critical pressure of 11GPa on application of pressure via diamond anvil cell (DAC) \cite{bise-press13}. Pressure induced structural phase transitions have been experimentally observed at very high pressures \cite{Yu2015}. The experiments indicate that a progressive structural evolution occurs from an ambient rhombohedral phase (Space group (SG): R$\bar{3}$m) to monoclinic phase (SG: C2/m) at 36GPa and eventually to a high pressure body-centered tetragonal phase (SG: I4/mmm) at 81GPa on application of pressure via DAC.  A pressure induced transition to a topological phase has been found in Bi$_2$S$_3$ at a pressure of 5.3GPa exerted via DAC \cite{bis}.

In some of topological materials like TaAs single crystals the valance band (VB) and the conduction band (CB) intersects at two points at $\pm$ k near the Fermi energy. If band dispersion near the crossing point is linear due to relativistic nature of fermions and the system preserves the time-reversal and inversion symmetry, then the system may be characterised as  a Weyl semimetal (WSM) \cite{weyl_armitage, weyl_felser}.  In WSMs non-orthogonal magnetic and electric fields results in a novel observation of chiral anomaly. This results in the chiral-magnetic effect which is the observation of an unconventional negative longitudinal magnetoresistance \cite{Arnold2016}.

In this article we investigate the effects of hydrostatic pressure (HP), as well as doping with a rare earth element on bulk phase of Bi$_2$Se$_3$, a probable 3D TI, from the perspective of ab-initio density functional theory (DFT) based calculations. The material shows a electronic topological transition from a small bulk band gap insulator at low HP, to a gapless dirac state at a critical pressure 7.9 GPa, and to a WSM beyond the critical pressure. To the best of our knowledge this is the first prediction of a WSM state arising due to the application of pressure in Bi$_2$Se$_3$ family of materials. We note that Bi$_2$Se$_3$ undergoes a transition from a layered quasi 2D van der Waals crystal to 3D topological material on applying HP $P>$ 7.9 GPa.
We also study the effect of doping of Bi$_2$Se$_3$ with rare earth elements.  While intercalating Gd between the QLs shows a broad bandwidth metallic ground state, with a time reversal symmetry broken Weyl like feature in the band structure below the Fermi energy, substituting Bi with Gd shows an increase in band gap and an insulating state. Thus we propose that a tunable topological transition to Weyl states may be driven in bulk Bi$_2$Se$_3$ by both application of hydrostatic pressure as well as intercalating with rare earth elements, to attain the exciting and elusive Weyl like states.

The paper is divided into five sections, we describe the computational method in section II. The results are divided into two major sections. In section III we describe the effect of pressure on structural and electronic properties. The effect of doping is studied in section IV. The summary and conclusion drawn from the results are given in section V.

\section{Computational details}
Our first-principles calculations were carried out in the plane wave basis as implemented in the Vienna Ab-initio Simulation Package (VASP) \cite{vasp} with projector-augmented wave (PAW) potential \cite{paw}. The exchange-correlation functional used in the calculations is the generalized gradient approximation (GGA) implemented following the Perdew-Burke-Ernzerhof \cite{pbe}  prescription. Local correlations are taken into account wherever necessary with the energy correction within the framework of GGA+U formalism primarily for dopant Gd atoms, with values of $U=6eV, J=1eV$. For ionic relaxations, internal positions of the atoms are allowed to relax until the forces became less than 0.005 eV/A$^0$. Energy cutoff used for calculations is 500 eV, and $6\times 6\times 4$ Monkhorst-Pack k-points mesh  provide a good convergence of the total energy in self-consistent field calculations.  The spin-orbit coupling (SOC) in Bi atoms is treated as a perturbative non-self consistent correction which is better suited for topological materials \cite{Pakdel2018}. In order to study the effect of hydrostatic pressure, calculations are done by first changing the volume of the unit cell isotropically and then relaxing the ionic positions for each of the modified volume. The van der Waals corrections (\'a la DFT-D3 method of Grimme with Becke-Johnson damping) were also included in our VASP simulations. The phonon spectrum was calculated based on the density functional perturbation theory (DFPT) as implemented in the VASP package. A 2$\times$ 2$\times$ 1 supercell and a $\Gamma$ centered 2$\times$ 2 $\times$ 1 Monkhorst-Pack $k$ mesh  were  used.  The  phonon  frequencies were calculated using the Phonopy code \cite{phonopy} 

\section{Effect of Hydrostatic Pressure $P$}
In this section we discuss the effects of the application of hydrostatic pressure (HP) P on bulk band structure of  Bi$_2$Se$_3$. The HP induced structural transition, and changes in lattice parameters are analysed in first subsection, and a systematic study of the effect of structural transition on the band structure  properties of Bi$_2$Se$_3$ are provided in the next subsection.
Generally, pressure pushes the atoms closer to each other, and leads to enhancement in effective hybridization of orbitals which result in reduction of band gap. In large $P$ limit structural transition is also a possibility. In this section we study the structural behaviour of Bi$_2$Se$_3$ first, and then analyse the effect of pressure induced structural transitions on the electronic properties. 

\subsection{Change in crystal structure}
 We first discuss the basic structural details of the material at $P=0$GPa, and thereafter, the structural changes are analysed at various pressures.
\begin{figure}[h]
\begin{center}
\includegraphics[width=\linewidth]{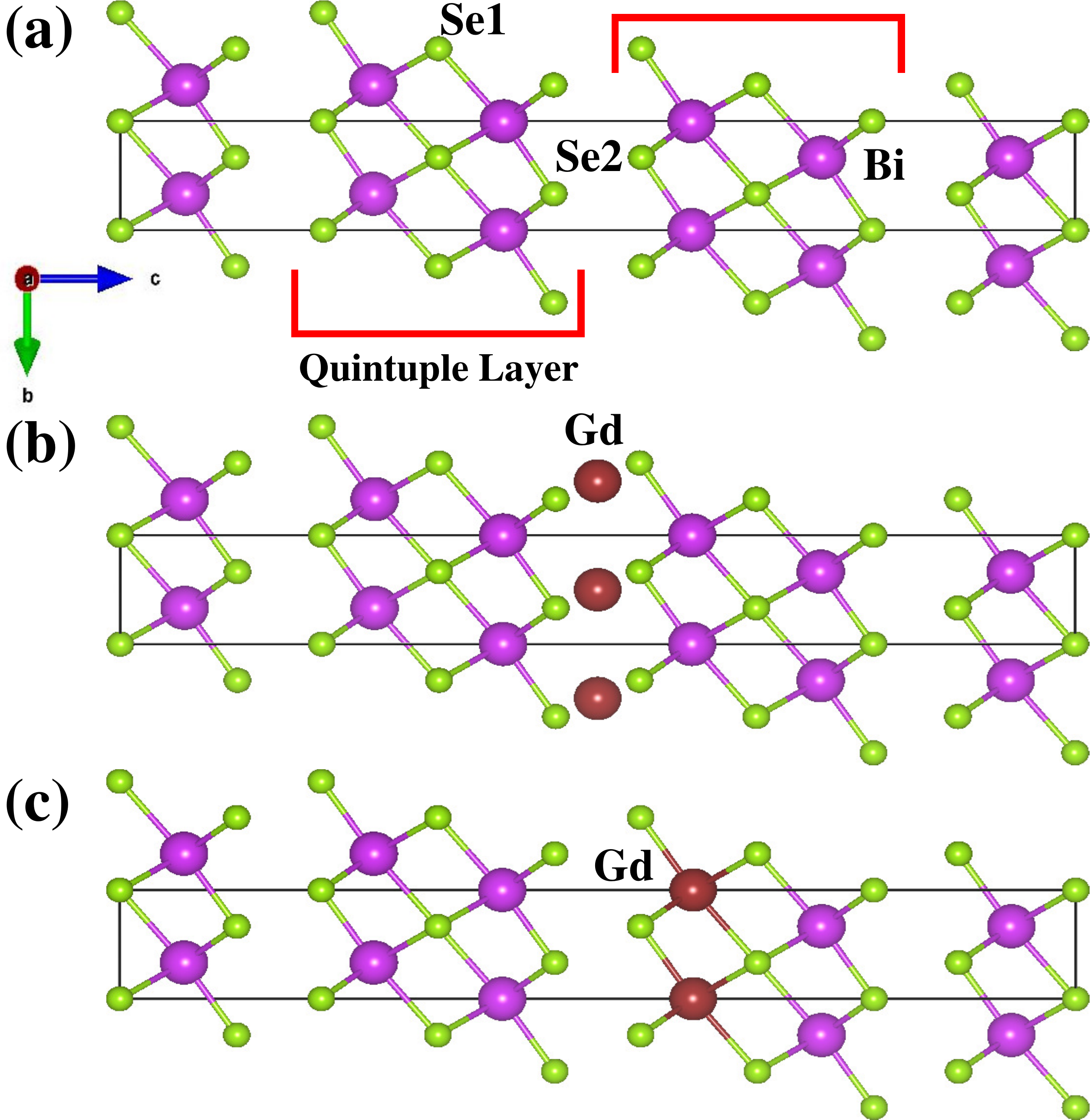}
\caption{(Color online) Figure showing the crystal structure of Bi$_2$Se$_3$. The violet spheres denote the Bi ions and the green spheres denote the Se ions. As can be seen from the structure Bi$_2$Se$_3$ forms quintuple layers. The top panel (a) shows pure Bi$_2$Se$_3$, the middle panel (b) shows Bi$_2$Se$_3$ with Gd intercalated between the quintuple layers, and the bottom panel (c) shows Bi substituted with Gd in Bi$_2$Se$_3$.}
\label{struct}
\end{center}
\end{figure}
Bi$_2$Se$_3$ has a hexagonal symmetry with space group R$\bar{3}$m, and with lattice parameters a=b=4.142 $\AA$, c=28.637 $\AA$ and lattice angles $\alpha=\beta=90^{\circ}$, $\gamma=120^{\circ}$. Bi$_2$Se$_3$ forms quintuple layers (QL) within the hexagonal unit cell as seen from Fig \ref{struct} a. The crystal structure along the c-axis direction consists of QLs of two Bi layers sandwiched between three Se layers, Se$_1$-Bi-Se$_2$-Bi-Se$_1$, where the subscript indicates that the two Se atoms are in-equivalent by symmetry (as shown in Fig.\ref{struct}). The atoms within each QL are chemically bonded, but the QLs are weakly bonded through van der Waals interaction. We show in Fig. 2(b) the electron localisation function (ELF) at 0 Gpa. It is seen that within the QLs there is covalent bonding by electron sharing between Bi and Se. No such electron overlap or sharing is seen between the layers, which are loosely connected by van der Waals interaction, and shows a clear gap in the ELF between the QLs. Bi$_2$Se$_3$ slabs consist of an integer number of QLs.

To understand the structural changes with pressure $P$, $c/a$ ratio and distance between Se-Se atoms sitting at two nearest QL $d_{Se-Se}$  are studied,  and we note that $c/a$ ratio decreases with $P$ up to $P= P_c=7.9$ GPa, and then it increases on increment of $P$ as shown in Fig. \ref{trans} a. Variation of lattice parameter a, b and c and c/a with $P$ are provided in  Table I in the Appendix. 
The distance between the two $Se$ atoms from two nearest QL $d_{Se-Se}$ continuously decreases with $P$, and at  $P=P_c$ the Se-Se distance decreases to below 2.17 $\AA$ which is less than bond length of diatomic $Se_2$ \cite{hub-her}. Therefore for $P>P_c$  Bi$_2$Se$_3$ behaves like three dimensional (3D) structure rather a layered structure, as demonstrated in Fig \ref{trans} c.
The bond distances from Bi to Se1, $d_{Bi-Se1}$, and to Se2, $d_{Bi-Se2}$,  decrease on application of pressure. The bond angles $\angle$Se1-Bi-Se1 and $\angle$Se2-Bi-Se2 also decrease with increase of $P$, however due to bending of the bonds, $\angle$Se1-Bi-Se2 increases with increase of $P$ as shown in Table II in the appendix.

The structural change from a quasi 2D layered van der Waals crystal to a 3D crystal can be seen in the ELF plotted in Fig. 2(c), unlike the $P=0$ Gpa case, there is an electron sharing or overlap of orbitals between the QLs. Therefore we can confirm that the structural change leads to the electronic topological transition.
We also note that our  calculation shows no significant effect of the vdW correction (\'a la DFT-D3 method of Grimme with Becke-Johnson damping) implemented in VASP.

We studied the dynamical stability of the Bi$_2$Se$_3$ by performing phonon calculations and analyzing the phonon spectrum. In absence of pressure, Bi$_2$Se$_3$ has small negative frequencies $\omega \le -0.5$ THz which is consistent with the earlier studies~\cite{bise-phonon}. The negative frequencies are related to the negative stress tensor of the optimized system. The lattice parameters being optimized using GGA to reduce the internal force, overestimates the lattice constant. As a result, the optimized structure has a negative stress tensor that leads to negative frequencies. The phonon modes along z-direction are sensitive to stress because of the weak vdW force along that direction. External pressure favors positive frequencies by decreasing the lattice constant. This is shown in the Fig. 8 in the appendix which depicts that the contribution from the negative frequencies to the phonon density of states decreases with increasing pressure and, at $P=33.4$ GPa, the negative frequencies are negligibly small. Thus higher pressure leads to greater stability in the system, which has been predicted in existing literature. \cite{bise-phonon}. 

\begin{figure}
\includegraphics[width=\columnwidth]{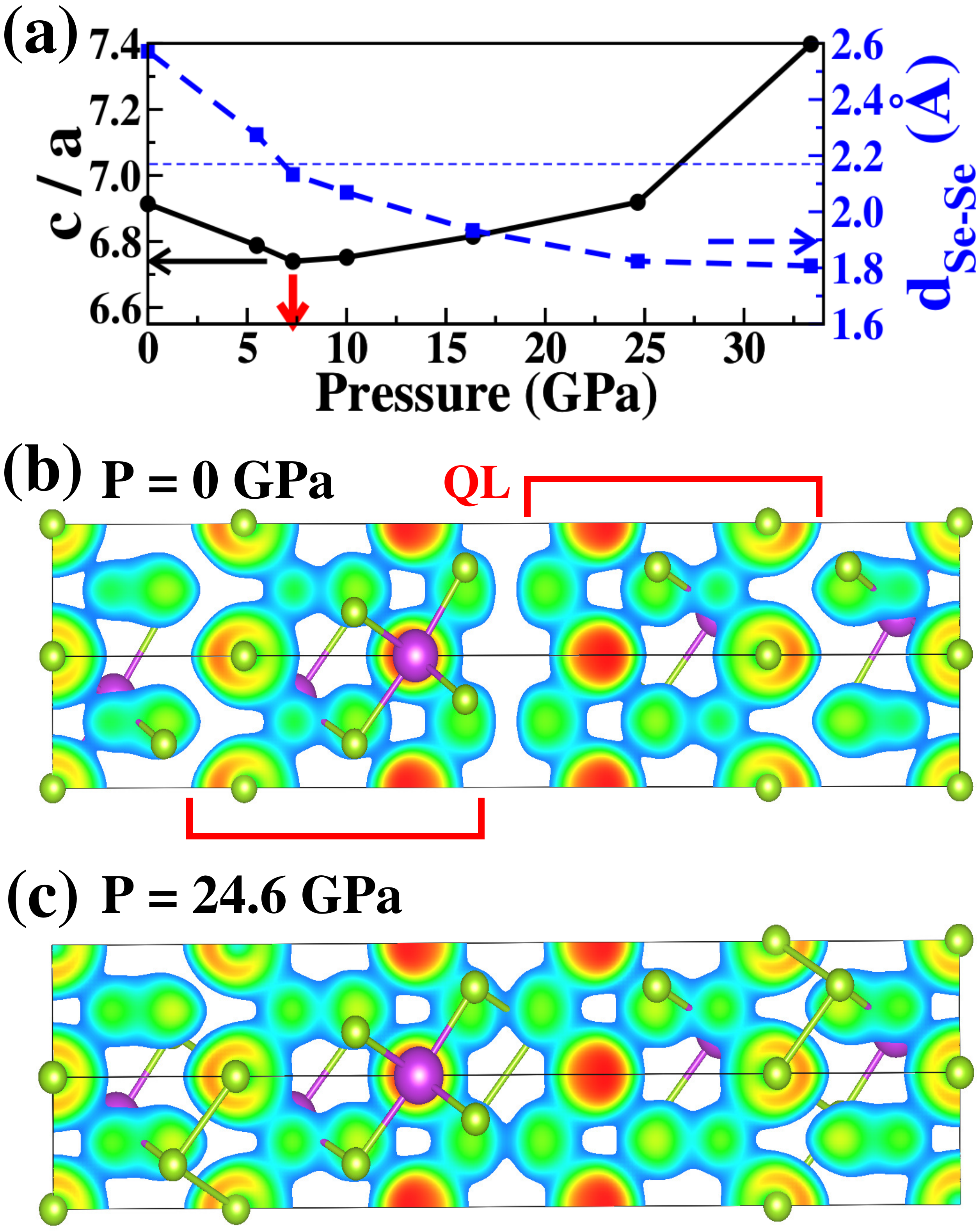}
\caption{(Color online) Figure showing structural transition on application of Hydrostatic Pressure. Top panel (a) shows the change in $c/a$ ratio and the distance between the quintuple layers ($d_{Se-Se}$ with pressure. The blue dashed line marks the bond length of diatomic $Se_2$.
A red arrow at $P = 7.3$ GPa marks the point of structural transition from a layered 2D structure to a 3D structure. The bottom panel (b) and (c) shows electron localisation functions at two different pressures  at 0 Gpa and 24.6 Gpa  before and after the critical pressure for structural transition from a layered vdW crystal like 2D structure to a 3D like structure with inter-layer bonding.}
\label{trans}
\end{figure}

\subsection{Change in electronic structure}
\begin{figure*}
\begin{center}
\includegraphics[width=1.75\columnwidth]{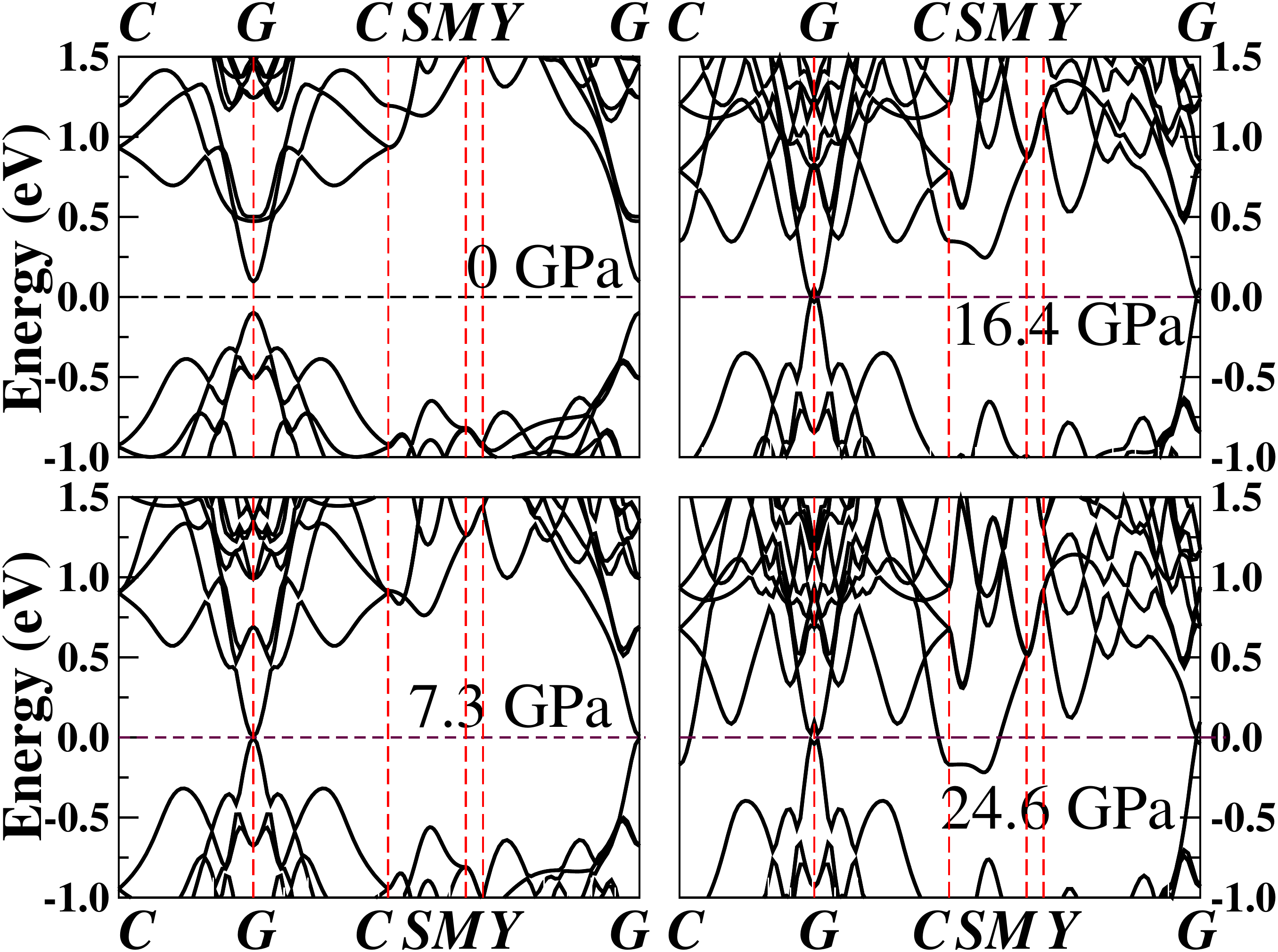}
\caption{(Color online) Figure showing the variations in total band structures with the exertion of pressure. We observe the bands at $\Gamma$ point denoted by G approaching each other and finally interpenetrating beyond the critical pressure to show the Weyl semimetal states.}
\label{band-press}
\end{center}
\end{figure*}
We study the change in electronic structure of Bi$_2$Se$_3$ with application of hydrostatic pressure as shown in Fig. \ref{band-press}. The band structure at zero applied pressure has a band gap of 0.3 eV at the $\Gamma$ point and both Bi and Se have partially filled $p$-orbitals which participate to form energy bands as seen from the projected DOS in Fig. \ref{dos-press}. The valence bands are primarily Se-$p$ bands, whereas, the conduction bands are Bi-$p$ bands as shown in Fig. \ref{band-press}. Based on symmetry analysis we find that the $p$ levels on the Bi and Se1 are split by the crystal symmetry into $p_z$ and ( $p_x$ , $p_y$ ) at the $\Gamma$ point. The band gap is formed between the bonding and anti-bonding states resulting from the hybridization of $p_z$ orbitals on the Bi and Se1 sites. Considering surface calculations the band structure of Bi$_2$Se$_3$ shows the presence of surface states at the $\Gamma$ point which is a typical signature of topological insulators at $P=0$ (Not shown here. This has already been shown in several papers (Fig.4 of Ref.~\onlinecite{zhang}, Fig.1 of Ref~\onlinecite{oleg2010}) ).

\begin{figure}
    \centering
    \includegraphics[width=\columnwidth]{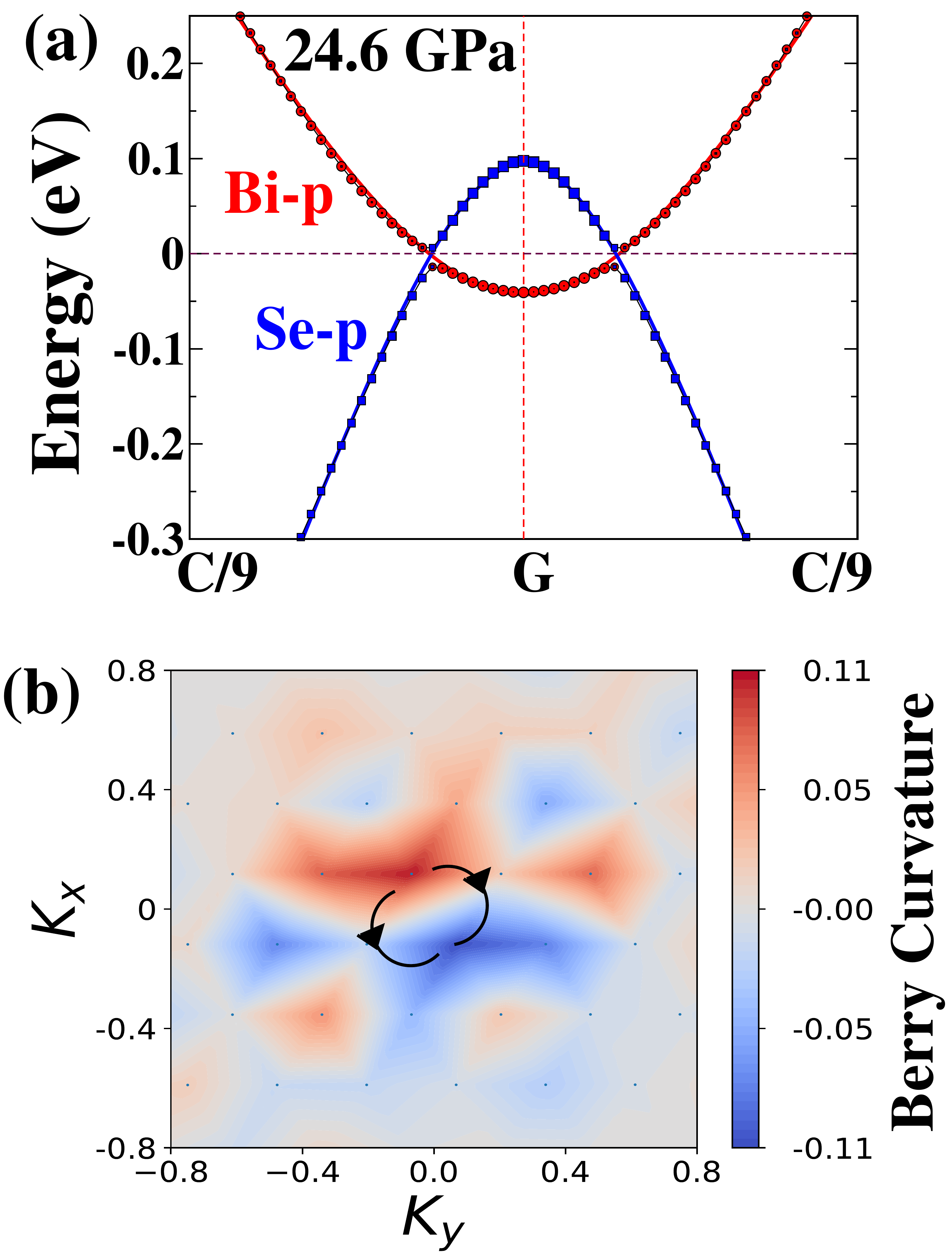}
    \caption{(Color online) Figure showing the emergence of Weyl states in Bi$_2$Se$_3$. Panel (a) shows the atom resolved band structure showing the Weyl points and the band inversion around the Fermi energy at a Pressure of 24.6GPa. Panel (a) also shows the bands near the band dispersion fitted to Eqns \ref{valence} and \ref{conduction} corresponding to band dispersions for Weyl semimetals. The fitting is shown with bold lines corresponding to the atom resolved band structure. Panel (b) shows the Plot of Berry Curvature which shows the distinctive case of Weyl semimetals.}
    \label{weyl}
\end{figure}

\begin{figure}
\begin{center}
\includegraphics[width=\linewidth]{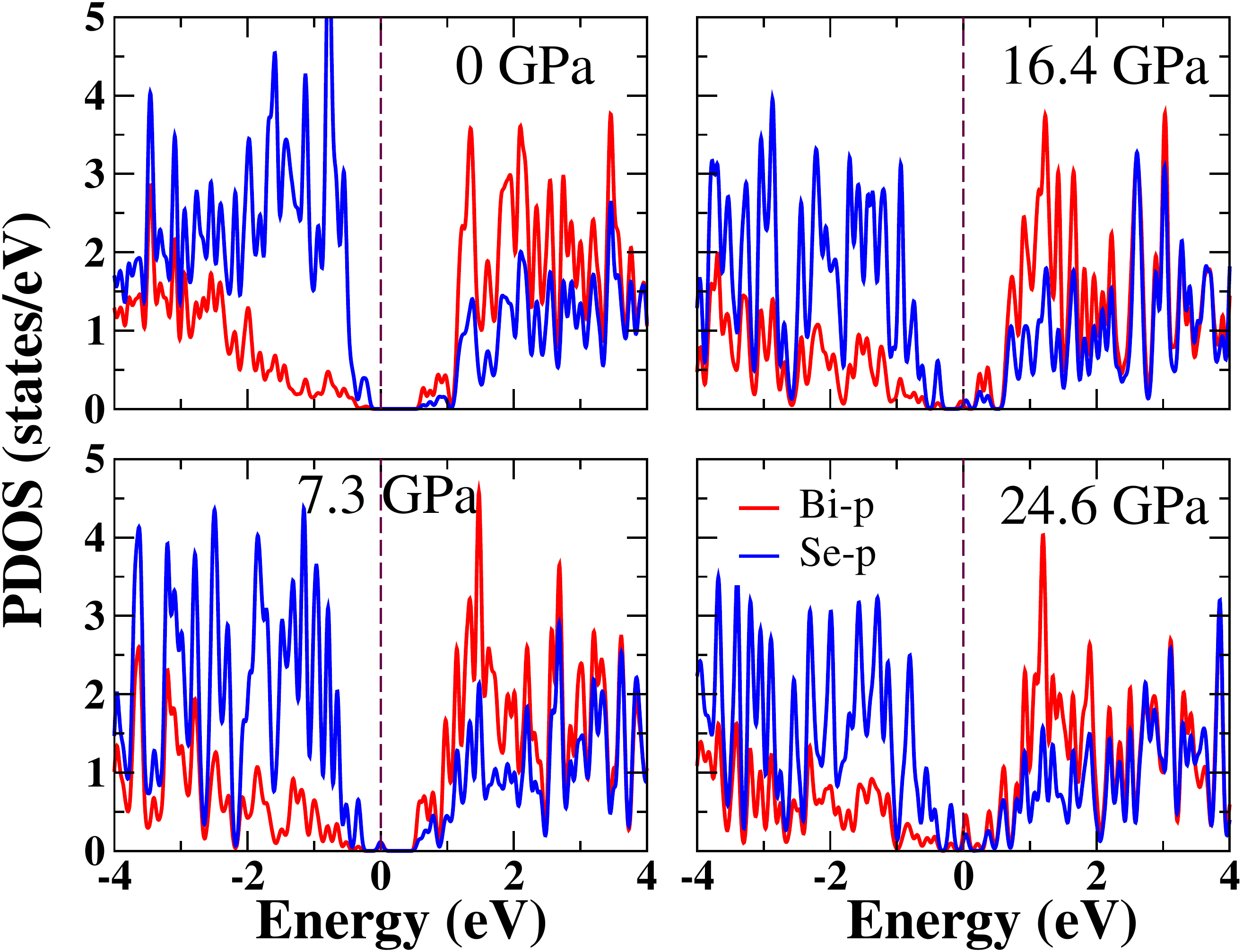}
\caption{(Color online) Figure showing the variations in Projected density of states (PDOS) with the exertion of pressure. The energy is scaled with respect to Fermi energy. The red lines represent Bi $p$ contribution while the blue lines represent Se $p$ contribution.}
\label{dos-press}
\end{center}
\end{figure}

We apply HP $P$ systematically on the system and notice that the band gap decrease with $P$ and it is 0.05 eV and  0.009eV at $P =   5.5$  and  $7.3$ GPa respectively. The band gap vanishes completely at $P_c = 7.9$ GPa,  and this gapless state can be correlated to the structural phase transition at $P >7.9$GPa. Surprisingly, the CB and VB moves towards the fermi-energy with increasing $P$ as shown in Fig.\ref{band-press} before the critical pressure $P_c$ at which VB and CB meet each other and the band gap vanishes. Whereas, for $P>P_c$ these two bands crosses at two points $\pm$ k points around the $\Gamma$ point at the fermi-energy. There is no spin-splitting at these crossing points and these crossing points are possible signatures of WSM as shown in Fig. \ref{weyl}.

In a Weyl semimetal other than CB  and VB coinciding within some energy window, the degeneracy is expected to be robust to small parametric perturbation. The double degeneracy may arise in presence of  time reversal T and inversion symmetry P or their combined PT presence in the system \cite{weyl_armitage} i.e. for inversion symmetry  $E_{n \sigma}(k)=E_{n \sigma}(-k)$ , for time reversal symmetry $E_{n \uparrow}(k)=E_ {n \downarrow} (k)$  and in combined PT symmetry $E_{n \uparrow}(k)=E_ {n \downarrow} (-k)$ conditions are satisfied.  These conditions are easily fulfilled in case of band inversion,  i.e., the two branches of a band undergo an accidental band crossing and give rise to Weyl points, and this is applicable in our case.  However the degeneracy of crossing point are preserved only in case special symmetry in the lattice, and the symmetry prevents the repulsion of degenerate points to keep the four fold degeneracy intact. Bi$_2$Se$_3$ has R$\bar{3}$m crystal symmetry and under pressure the distance between the QLs reduces and gives rise to three dimensional structure. For $P>7.3$ GPa, the Bi-p band and Se-p  band crosses at $K_x=\pm  0.045$  and $E_f=0$ at $P=24.6$ GPa as shown in Fig. \ref{weyl} a. The crossing point shown in the Fig. \ref{weyl} is along the $\Gamma$-C line, but similar crossing can be found along other $\Gamma$-K momentum axis. 

In absence of any external perturbation, the band dispersion near the Weyl point varies as $E(k)=\sqrt{m^2+v^2k^2}$  with  momentum $k$ where $m$ and $v$ are mass and velocity parameter. 
In our system the CB and  the VB are formed from two different orbitals which have different chemical potential. 
Therefore, the dispersion relation for VB and CB near the Weyl point can be fitted with  
\begin{equation}
   \frac {E +0.39} {0.9} = \sqrt{ 1700 (k-0.051)^2+1.4 (k-0.051)+0.149} 
   \label{valence}
\end{equation}
 and    
 \begin{equation}
     \frac {E -0.435} {0.89} = -\sqrt{ 4300 (k-0.051)^2+3.8 (k-0.051)+0.147}
     \label{conduction}
 \end{equation}
 as shown in Fig. \ref{weyl} a. We notice that crossing point is symmetric about the  $\Gamma$ point and spin up and down channel of the band is degenerate in absence of the  SOC.  Therefore, this system preserves the both time reversal and inversion symmetry. The crossing points acts as source or sink of the Berry curvature. The Barry curvature calculated using Vaspberry \cite{berry} shown in inset Fig. \ref{weyl} b. The Berry curvature have highest value  near one crossing point whereas it lowest value near the second crossing points, the direction of curvature is shown with arrow.
In Fig. \ref{dos-press} projected density of states (PDOS) are shown for four different pressures, and the Bi and Se $p$ bands are marked separately. The contribution of Bi-p band is higher near the $E_f$ in the CB than the Se-p band, whereas in VB Se-p band has higher contribution.  At low pressure  $P <7.3$ GPa there is no density of states at $E_f$, and PDOS at $E_f$ increases with $P$, large PDOS can be seen at $P=33.4$ GPa.  

\section{Effect of doping}
\begin{figure}[h!]
    \centering
    \includegraphics[width=\columnwidth]{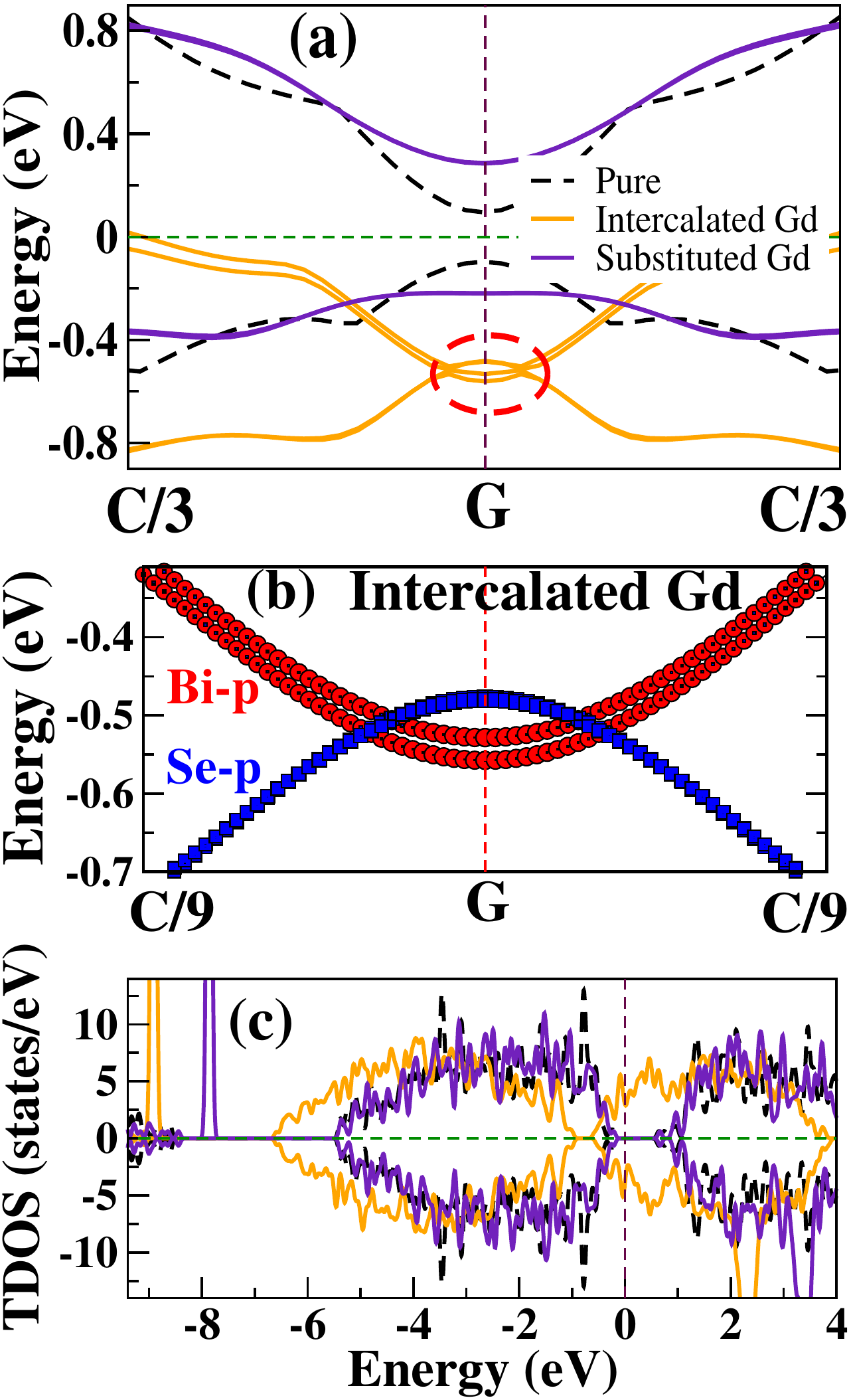}
    \caption{(Color online) Figure showing the effect of doping Gd both substitutionally and intercalating between the layers of Bi$_2$Se$_3$. The upper panel (a) shows the valence band maxima and conduction band minima for all three cases of pure Bi$_2$Se$_3$ (black dashed), Bi$_2$Se$_3$ intercalated with Gd between the layers (orange), and Bi substituted with Gd (violet) in Bi$_2$Se$_3$. The emergence of time reversal symmetry broken Weyl states, with the Weyl points and band inversion, in Bi$_2$Se$_3$ intercalated with Gd, also showing the magnetic splitting of the bands, is marked with the red dashed circle in this panel. The panel (b) shows the orbital resolved band structure for the case of intercalated Gd, proving the band inversion below the Fermi level.} The panel (c) shows the total DOS for the same cases. The energy is scaled with respect to the Fermi energy.
    \label{dope-es}
\end{figure}
In this section we discuss the effect of doping Bi$_2$Se$_3$ with a rare earth element Gd. The system can be doped either by the substituting Bi with Gd, or intercalating Gd between the QLs of Bi$_2$Se$_3$. Intercalating one Gd per unit cell gives rise to a doping fraction of 16.67\% whereas substituting one Bi with Gd per unit cell gives rise to a doping fraction of 20\%. To our surprise the two different methods of doping, with similar doping fractions, resulted in two completely different electronic ground states. The highest valence and the lowest conduction bands are shown in Fig. \ref{dope-es} a for pure, substituted  and intercalated Gd in Bi$_2$Se$_3$ system. On substituting Bi with Gd, there is an increase in the split between the valence band maxima and conduction band minima, and hence the direct band gap increases to 0.5eV. 
The intercalation of Gd shifts the entire band structure to a lower energy in such a way that we now have a partially occupied conduction band, and a fully occupied VB as shown in Fig. \ref{dope-es} (a), and therefore it is a wide band metal. The intercalated Gd induces rearrangement of energy bands which leads to band inversion below $E_f$, however, it is different from the pressure induced inversion. The band inversion below $E_f$ is shown in the circle in Fig. \ref{dope-es} (a). In this case the up and down spin band split due to internal magnetic field induced by Gd atoms. Therefore the resulting time reversal symmetry is not preserved. The upper band is contributed from the Bi-band whereas the lower band contributed by Se atoms. This is shown in the orbital resolved band structure for Gd intercalated system in Fig. 6(b). Thus we observe a time reversal symmetry broken Weyl state whose existence has been discussed in literature. \cite{time-rev}.

We note here that on carrying out intercalation of Gd at a lower doping fraction of 8.33\%, which is 1 Gd atom intercalated in a 2$\times$ 1$\times$ 1 supercell of undoped Bi$_2$Se$_3$, we find qualitatively same electronic structure, with very similar DOS and a similar band inversion below the Fermi energy.

The total DOS for three different cases pure, substituted and intercalated Bi$_2$Se$_3$  are  shown in Fig \ref{dope-es}(b), and we note that intercalation of Gd can lead to finite density of states at $E_f$ which leads to an insulator to metal transition. The large DOS comes from Gd $f$ orbitals which are half filled and have high spin splitting. The rest of the DOS has usual contributions from Bi and Se $p$ orbitals with some mixing with filled Gd $d$ orbitals. The contribution to DOS at the Fermi in case of Gd intercalated Bi$_2$Se$_3$ is from the Bi $p$ bands which is the partially filled conduction band. It is also to be noted that although the direct band gap changes in case of substitutional doping from pure Bi$_2$Se$_3$, the integrated band gap does not undergo any substantial change as observed from the total DOS.

\begin{figure}
    \centering
    \includegraphics[width=\columnwidth]{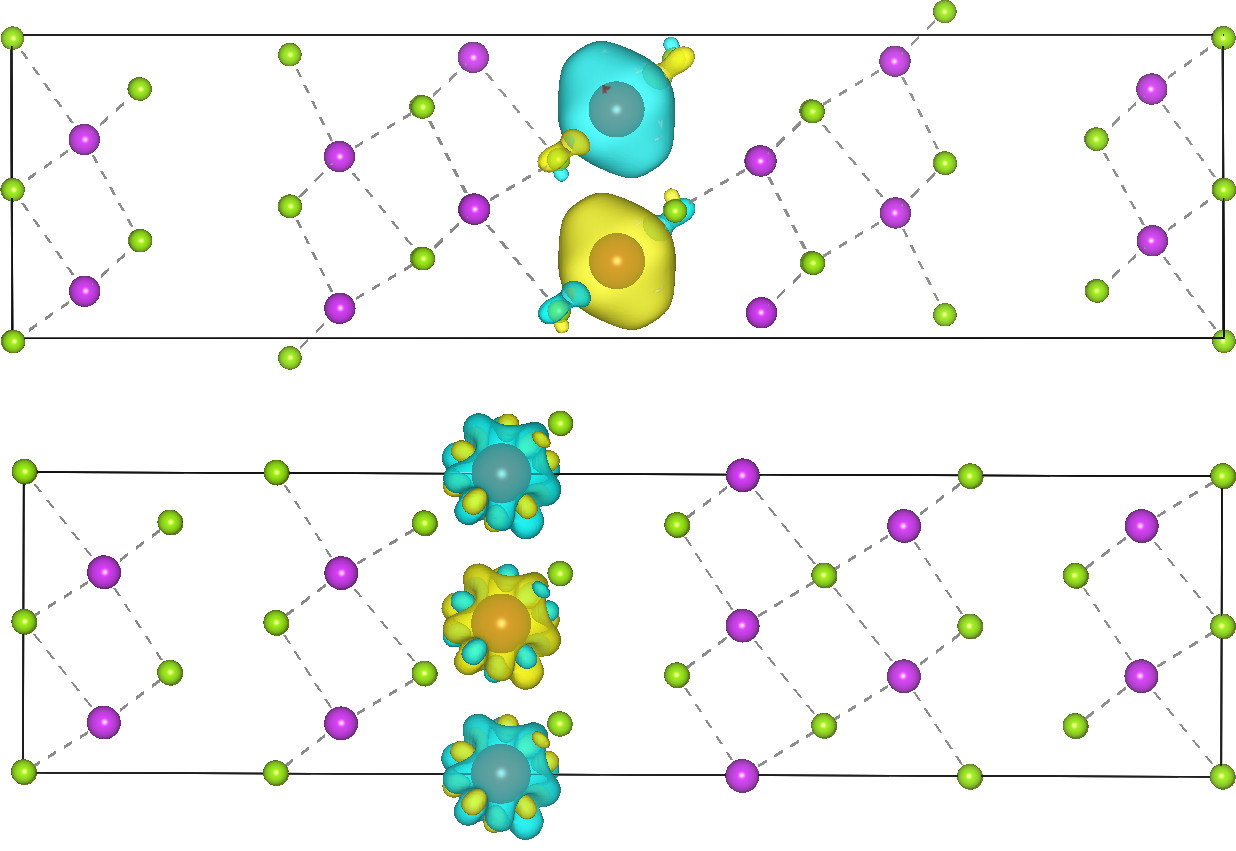}
    \caption{(Color online) Figure showing the magnetisation density in both Gd intercalated and substituted Bi$_2$Se$_3$. The top panel shows the case of Gd intercalation while the bottom panel shows that of Gd substitution. Yellow isosurfaces represents positive direction of magnetic moment while blue isosurfaces represents negative.}
    \label{dope-mag}
\end{figure}

The magnetic properties of the system under the influence of the dopant atom changes significantly. The substitutional doping gives rise to a AFM ordering consistent with earlier results \cite{kim2}. We find a similar AFM ordering between intercalated Gd atoms, as shown in Fig. \ref{dope-mag}. The large magnetic moment of $\sim 7 \mu_B$ on Gd which is consistent with literature \cite{kim2}. The large magnetic moment on Gd induces a small moment of $\sim 0.2 \mu_B$ on Se. The large magnetic moment may be because of larger exchange splitting compared to the crystal field splitting, in this case, the $f$-orbitals  likely to be occupied by electrons with parallel spin. The large local magnetic moments induces Zeeman splitting in the Bi and Se $p$ bands in both doping method, although intercalation of Gd has a much stronger splitting effect on the Bi and Se $p$ bands than substitutional doping as seen from Fig\ref{dope-mag}. Thus intercalation is more effective in inducing magnetism than previously reported methods of substitution. 

To estimate the spin exchange coupling in the solid state system, generally one considers at least two magnetic atoms per supercell. We consider a 2$\times$1$\times$1 supercell for this purpose. 
We consider the moments to interact via a simple nearest neighbour Ising model $E=J\times S_iS_j$, where S=7/2. We find a magnetic exchange J=9.6meV between the Gd atoms, which is mediated by the Se $p$ orbitals interacting anti-feromagnetically, as seen in from the plot of magnetisation density which is shown in Fig. \ref{dope-mag}. However, the orbital structures for the two different doping are quite different which might be related to the different conductivity behaviour. Although in both cases majority of magnetisation comes from the $f$ orbitals (in both cases a combination of $f(3x^2-y^2), f(xyz), f(yz^2)$ orbitals, there is a rather significant contribution to the total moment of 7$\mu_B$ as well from $d$ orbitals. In case of intercalation $d(x^2-y^2)$ has a major contribution while in case of substitution $d(xz)$ has a major contribution. This shows up in the magnetisation density plotted in Fig \ref{dope-mag}.

Thus our study on doping not only shows a tunable topological transition to Weyl like states depending on the method of doping but also shows induced magnetism in the system.

\section{Summary and Conclusion}
In our ab initio DFT study we note that bulk Bi$_2$Se$_3$ shows a tunable topological transition with the application of hydrostatic pressure and by doping  with rare earth elements. Bulk Bi$_2$Se$_3$, upon application of hydrostatic pressure, shows an electronic topological transition from a surface states driven topological insulator to a Weyl semimetal for $P>P_c$, and the transition can be associated with structural changes from a layered quasi 2D vdW material to a 3D material.  For $P>P_c$ Se-p band and Bi-p band shows band inversion and these two bands cross at two points, $\pm$ k points or Weyl points around the $\Gamma$ point at the Fermi-energy. There is no spin-splitting at these crossing points, and $E_{n \uparrow}(k)=E_ {n \downarrow} (-k)$. We also notice that crossing points shifts to higher momentum  $k$ for larger $P$, whereas  the DOS at the Fermi-energy  have finite value at large $P$. 

Our DFT calculations also suggest that a topological transition may also be achieved by doping, albeit depending on the type of doping. Intercalating a rare earth atom, Gd, between the QL leads to a metallic state with a large band width, and shows presence of Weyl points below the $E_f$; while substitution leads to a larger direct band gap compared to undoped Bi$_2$Se$_3$. However both type of doping induces anti-ferromagnetic ordering in the system. We also note an induced magnetism in the system owing to the large magnetic moment on the rare earth dopant Gd atom.

The anti-ferromagnetic metallic state is particularly important for spintronics based application which may be driven in this material. This anti-ferromagnetic metallic state arises in case of intercalation of Gd between the QLs, and could be of immense importance in spintronics based applications \cite{Smejkal2018}. Anti-ferromagnetic metals have primarily been seen as exotic electronic structure states. Our study opens up new application possibilities in this 3D topological material, and most importantly shows the emergence of Weyl semimetal state in the Bi$_2$Se$_3$ family of materials by application of pressure which may be easily verified experimentally. Similar experiments may also be carried out with other materials in this class like Sb$_2$Te$_3$, Bi$_2$Te$_3$ and Bi$_2$Se$_3$.

We hope our study motivates further theoretical and particularly experimental studies to explore the tunable topological transitions, particularly in the search for exotic Weyl semimetals, with associated magnetism in TIs both from the perspective of understanding of novel states and device applications as well.

\begin{acknowledgments}
The authors thank DST India for the funding and also computational facilities at S. N. Bose National Centre for Basic Sciences, Kolkata, provided by the Thematic Unit of Excellence on Computational Materials Science. MK thanks DST India for a Ramanujan Fellowship SR/S2/RJN-69/2012. HB thanks the Austrian Science Fund (FWF), for funding through START project Y746, and DST India for funding support during execution of the project. SKS thanks DST-INSPIRE for financial support. HB acknowledges useful discussions with Dr. Sudipta Kanungo and Dr. Oindrila Deb. MK thanks Dr. Sandip Chatterjee and Dr. T. Setti for useful discussion.
\end{acknowledgments}

\section*{DATA AVAILABILITY}
The data that support the findings of this study are available from the corresponding author
upon reasonable request


\appendix
\section*{Appendix}

\begin{table}[h]
\caption{Table showing change in lattice parameters with exertion of hydrostatic pressure} 
\begin{tabular}{|c|c|c|c|c|c|c|}
\hline 
P(GPa) & 0 & 5.5 & 10 & 16.4 & 24.6 & 33.4 \\ 
\hline 
a ($\AA$) & 4.142 & 4.084 & 4.008 & 3.913 & 3.810 & 3.658 \\ 
\hline 
b ($\AA$)& 4.142 & 4.084 & 4.008 & 3.913 & 3.810 &  3.658 \\ 
\hline 
c ($\AA$) & 28.637 & 27.726 & 27.062 & 26.667 & 26.365 & 27.067\\ 
\hline 
c/a & 6.914 & 6.788 & 6.752 & 6.816 & 6.919 & 7.398 \\ 
\hline 
\end{tabular} 
\end{table}

\begin{table}[h]
\caption{Table showing change in bond angle and bond length with exertion of hydrostatic pressure (Note:Se1 is at the edge of the QL and Se2 is inside QL for pure Bi$_2$se$_3$).} 
\begin{tabular}{|c|c|c|c|c|c|c|}
\hline 
P(GPa) & 0 & 5.5 & 10 & 16.4 & 24.6 & 33.4 \\ 
\hline 
$d_{Bi-Se1}$ ($\AA$) & 2.86 & 2.84 & 2.81 & 2.78 & 2.74 & 2.71 \\ 
\hline 
$d_{Bi-Se2}$ ($\AA$)& 3.07 & 3.03 & 2.98 & 2.92 & 2.87 &  2.82 \\ 
\hline 
$\angle$ Se1-Bi-Se1 & 92.61 & 91.91 & 90.82 & 89.41 & 88.09 & 84.0\\ 
\hline 
$\angle$Se2-Bi-Se2 & 84.98 & 84.74 & 84.6 & 83.89 & 83.27 & 80.75 \\ 
\hline 
$\angle$Se1-Bi-Se2 & 91.13 & 91.6 & 92.2 & 93.17 & 94.19 & 97.22 \\ 
\hline 
\end{tabular} 
\end{table}

\begin{figure}[h]
    \centering
    \includegraphics[width=\columnwidth]{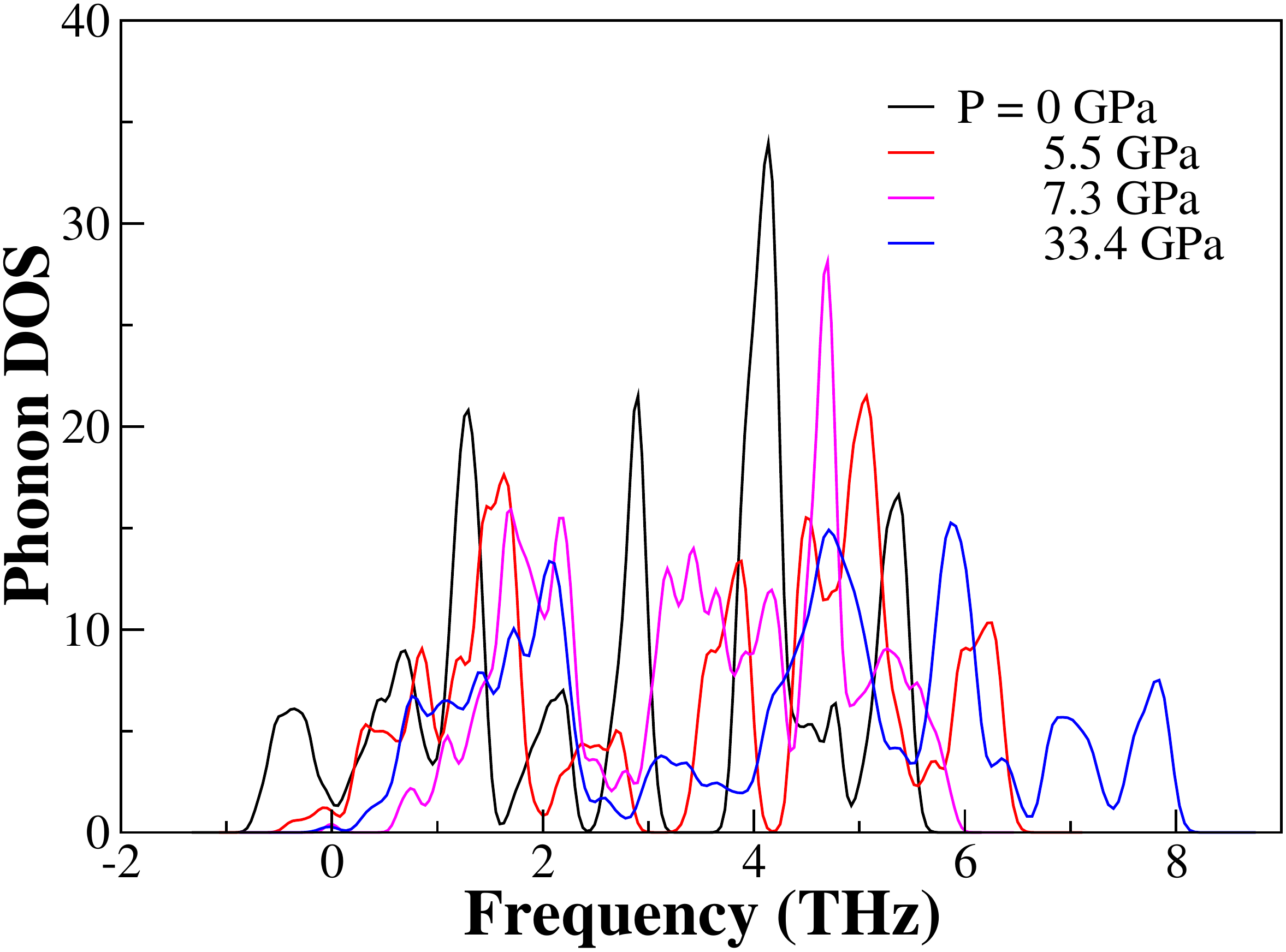}
    \caption{(Color online) Figure showing phonon DOS at different HP. In absence of pressure, phonon DOS is shown by the black line which has small but significant contribution from the negative phonon frequencies. It decreases with increasing HP. At $P = 33.4$ GPa, the contribution becomes negligible and thus brings greater stability to the system. }
    \label{phonon_dos}
\end{figure}

\bibliography{main}

\end{document}